\pdfoutput=1
\documentclass[aps,prd,reprint,twocolumn,showpacs,superscriptaddress,nofootinbib]{revtex4-1}  
\usepackage{tabularx}
\makeatletter
\def\hlinewd#1{%
\noalign{\ifnum0=`}\fi\hrule \@height #1 %
\futurelet\reserved@a\@xhline}
\makeatother
\usepackage{scrextend}
\usepackage{amsmath}
\usepackage{amssymb}
\usepackage{epsfig}
\usepackage{graphicx}
\usepackage{hyperref}
\usepackage{dcolumn}   
\usepackage{slashed}
\usepackage{color}
\usepackage{rotating}
\usepackage[margin=0.9in,a4paper]{geometry}
\usepackage[table,xcdraw,dvipsnames]{xcolor}
\usepackage[utf8]{inputenc}
\usepackage{colortbl}
\usepackage[normalem]{ulem}
\usepackage{mathrsfs} 
\usepackage[version=4]{mhchem}
\usepackage{cancel}

\definecolor{nicered}{rgb}{0.7,0.1,0.1}
\definecolor{nicegreen}{rgb}{0.1,0.5,0.1}
\definecolor{red}{rgb}{1.0, 0, 0}

\newcommand{\bdm}{\begin{displaymath}}
\newcommand{\edm}{\end{displaymath}}
\newcommand{\bea}{\begin{eqnarray}}
\newcommand{\eea}{\end{eqnarray}}

\def\be{\begin{equation}}
\def\ee{\end{equation}}

\definecolor{nicered}{rgb}{0.7,0.1,0.1}
\definecolor{nicegreen}{rgb}{0.1,0.5,0.1}
\definecolor{red}{rgb}{1.0, 0, 0}
\definecolor{niceblue}{rgb}{0,0,0.8}
\definecolor{red}{rgb}{1.0, 0, 0}
\hypersetup{colorlinks,citecolor= nicegreen,linkcolor= nicered,urlcolor=nicered}

\def\eq#1{{Eq.~(\ref{#1})}}

\def\fig#1{{Fig.~\ref{#1}}}

\def\sect#1{{Sect.~\ref{#1}}}

\def\gsim{\raise0.3ex\hbox{$\;>$\kern-0.75em\raise-1.1ex\hbox{$\sim\;$}}}
\def\lsim{\raise0.3ex\hbox{$\;<$\kern-0.75em\raise-1.1ex\hbox{$\sim\;$}}}

\def\mb[#1]{\mathbf{#1}}

\definecolor{LightCyan}{rgb}{0.88,1,1}
\definecolor{piggypink}{rgb}{0.99, 0.87, 0.9}
\definecolor{applegreen}{rgb}{0.55, 0.71, 0.0}
\definecolor{darkpastelgreen}{rgb}{0.01, 0.75, 0.24}
\definecolor{green-yellow}{rgb}{0.68, 1.0, 0.18}

\newcommand{\beq}{\begin{equation}}
\newcommand{\eeq}{\end{equation}}
\newcommand{\beqa}{\begin{eqnarray}}
\newcommand{\eeqa}{\end{eqnarray}}

\begin{document}



\title{RadioAxion results on the search for axion dark matter under Gran Sasso}

\author{Carlo Broggini}
\email{carlo.broggini@pd.infn.it}
\affiliation{\small \it Istituto Nazionale di Fisica Nucleare, Sezione di Padova, Via F.~Marzolo 8, 35131 Padova, Italy}

\author{Giuseppe Di Carlo}
\email{giuseppe.dicarlo@lngs.infn.it}
\affiliation{\small \it Istituto Nazionale di Fisica Nucleare, Laboratori Nazionali del Gran Sasso, 67100 Assergi (AQ), Italy}

\author{Luca Di Luzio}
\email{luca.diluzio@pd.infn.it}
\affiliation{\small \it Istituto Nazionale di Fisica Nucleare, Sezione di Padova, Via F.~Marzolo 8, 35131 Padova, Italy}

\author{Denise Piatti}
\email{denise.piatti@pd.infn.it}
\affiliation{\small \it Istituto Nazionale di Fisica Nucleare, Sezione di Padova, Via F.~Marzolo 8, 35131 Padova, Italy}

\author{Claudio Toni}
\email{claudio.toni@lapth.cnrs.fr}
\affiliation{\small \it LAPTh, Université Savoie Mont-Blanc et CNRS, 74941 Annecy, France}

\begin{abstract}
\noindent
We report first results from RadioAxion, an underground experiment searching for axion dark matter through periodic modulations of radioisotope decays. We monitor the $\alpha$ decay of $\ce{^{241}Am}$ via its $59.5$~keV $\gamma$ line using a NaI detector installed at the Gran Sasso Laboratory, where cosmic-ray-induced systematics are strongly suppressed. We present the measured spectra and the corresponding time-series analysis. No evidence for a periodic modulation is observed.
From these data we derive constraints on the 
axion decay constant
in the axion mass range from $10^{-21}$ to $10^{-9}$~eV.
\end{abstract}

\maketitle


\section{Introduction}
\label{sec:intro}

The search for time variations in radioisotope decay rates has a long history, dating back to the birth of radioactivity science. Already in her Ph.D.\ thesis, Marie Curie reported on an experiment comparing the radioactivity of uranium at midday and midnight, finding no statistically significant difference~\cite{MadameCurie}. In modern times, this subject has re-emerged: several studies have reported per-mille modulations in the decay constants of various nuclei, typically with annual periods but occasionally also monthly or daily ones (see \cite{McDuffie:2020uuv} and references therein). Conversely, other dedicated investigations found no evidence for such effects~\cite{Pomme1,Pomme2,Pomme3}. In particular, underground $\gamma$-spectroscopy measurements at the Gran Sasso Laboratory excluded modulations in several isotopes with amplitudes above a few parts in $10^{5}$ over periods ranging from a few hours to one year~\cite{Bellotti:2012if,Bellotti:2015toa,Bellotti:2013bka,Bellotti:2018jzd}. A central difficulty is that any genuine signal must be disentangled from environmental systematics, especially those associated with the cosmic-ray flux and its seasonal variation at the few-percent level. This makes underground operation crucial: the rock overburden suppresses muon- and neutron-induced backgrounds by several orders of magnitude, strongly reducing both cosmogenic rates and their seasonal modulations.

A well-motivated particle-physics origin for a genuine periodic modulation is axion dark matter. The QCD axion provides a compelling pathway beyond the Standard Model, simultaneously solving the strong CP problem~\cite{Peccei:1977hh,Peccei:1977ur,Weinberg:1977ma,Wilczek:1977pj} and offering a well-motivated dark-matter candidate~\cite{Dine:1982ah,Abbott:1982af,Preskill:1982cy}. 
This possibility remains theoretically well motivated and phenomenologically viable, although the axion parameter space depends on the cosmological history and is increasingly constrained by astrophysical, cosmological, and laboratory searches.
In the standard misalignment mechanism, the axion field is initially displaced from the minimum of its potential. 
When the Hubble expansion rate drops below the axion mass, the field begins to oscillate coherently around the minimum; for non-relativistic axion dark matter this corresponds to a classical background field, $a(t)\simeq a_0\cos(m_a t)$, up to corrections due to the small axion velocity dispersion. 
The oscillation frequency is therefore set by the axion Compton frequency, $\omega_a\simeq m_a$, or $\nu_a^{\rm}=m_a/(2\pi)$ in natural units.
The oscillating axion field 
can therefore induce small, periodic shifts of low-energy hadronic and nuclear quantities. In particular, the model-independent axion coupling to gluons promotes the QCD topological angle to a time-dependent quantity, $\theta\to\theta(t)$, implying oscillatory hadronic observables such as the neutron electric dipole moment~\cite{Graham:2013gfa,Budker:2013hfa,Stadnik:2013raa}, with complementary detection strategies discussed e.g.~in 
\cite{Irastorza:2018dyq,DiLuzio:2020wdo,Sikivie:2020zpn}. The same mechanism can also lead to a time modulation of nuclear decay rates, through the $\theta$ dependence of nuclear properties, relevant for $\beta$ and $\alpha$ decays.

Recently, the authors of Ref.~\cite{Zhang:2023lem} proposed to search for time variations in radioisotope decay rates by exploiting the $\theta$ dependence of $\beta$ decay developed in Ref.~\cite{Lee:2020tmi}, and used tritium data from the European Commission's Joint Research Centre~\cite{Pomme2} to set bounds on the axion-gluon coupling. Complementarily, in Ref.~\cite{Broggini:2024udi} we developed a theoretical framework for the $\theta$ dependence of $\alpha$-decay half-lives, showing that an oscillating axion background can induce a periodic modulation of $\alpha$ radioactivity. 
This provides the conceptual foundation for RadioAxion, an experiment dedicated to monitoring the $\alpha$ decay of $\ce{^{241}Am}$ through its characteristic radiative channel,
$\ce{^{241}Am}\to \ce{^{237}Np}+\alpha+\gamma(59.5~\text{keV})$,
so that the $59.5$~keV $\gamma$ line can be continuously recorded as a proxy for the decay rate. Similarly, in Ref.~\cite{Alda:2024xxa} we investigated the time modulation of weak nuclear decays (electron capture and $\beta$ decay) as a probe of axion dark matter. By developing a dedicated framework for the $\theta$ dependence of weak nuclear decays and recasting underground Gran Sasso data on $\ce{^{40}K}$ \cite{Bellotti:2018jzd} and $\ce{^{137}Cs}$ \cite{Bellotti:2012if}, we derived constraints on the axion decay constant in the axion-mass window from a few $10^{-23}$~eV up to $10^{-19}$~eV.

\begin{figure*}[t!]
\centerline{
\includegraphics[width=0.85\textwidth,angle=0]{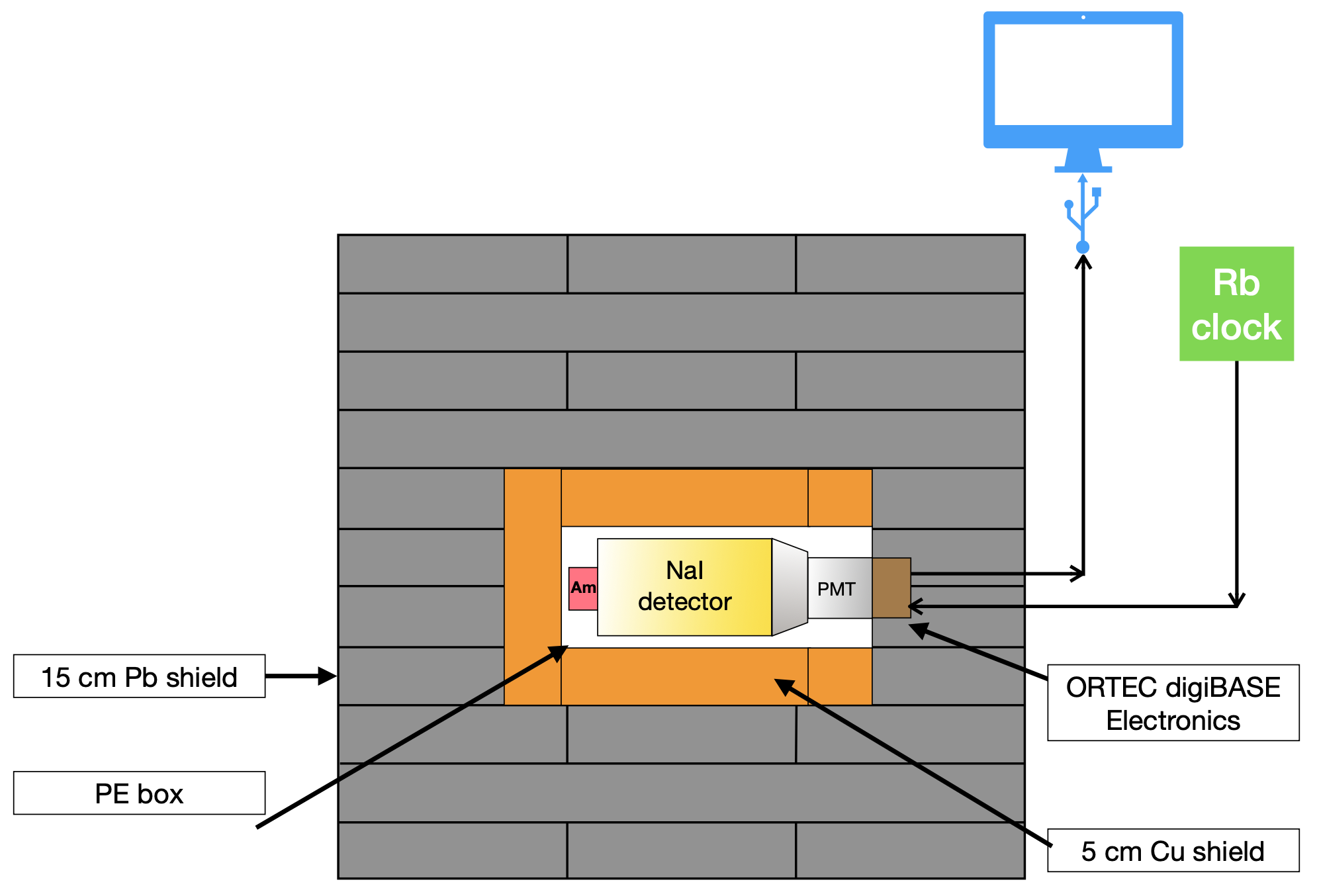}
\vspace{0.3cm}
}
\caption{Schematic (not to scale) of the RadioAxion setup installed at the Gran Sasso Laboratory. A $3''\times 3''$ NaI crystal views the $^{241}$Am source mounted in front of the detector; the scintillation light is read out by a PMT coupled to an ORTEC digiBASE unit.  
The detector assembly is enclosed in a polyethylene housing and surrounded by passive shielding consisting of $5$~cm oxygen-free high-conductivity copper and $15$~cm lead. Events above threshold are recorded in list mode and sent to the acquisition PC, while timing is stabilized by injecting a 1~Hz reference derived from a rubidium frequency standard (Rb) into the digiBASE.}
\label{fig:setup_am}
\end{figure*}

In this work, we report results from the first data-taking campaign of the RadioAxion experiment at the underground Gran Sasso Laboratory. 
Data were collected in two configurations: 266 one-day runs targeting oscillation frequencies from $1~\text{Hz}$ to $0.5~\text{MHz}$, and a single continuous run of 69 days sensitive to frequencies from $6.6\times10^{-7}~\text{Hz}$ up to $0.5~\text{Hz}$ (periods from $2$~s to $18$~days), enabling us to set constraints on the axion-gluon coupling 
(or, equivalently, on the axion decay constant)
over a broad axion-mass range, between $10^{-21}$ and $10^{-9}$~eV.

The paper is organized as follows. In \sect{sec:setup} we describe the experimental apparatus and the data acquisition. 
In \sect{sec:data} we present the data set and the analysis strategy. 
The physical interpretation in terms of constraints on axion dark matter is discussed in \sect{sec:axionDM}. 
Future developments are outlined in \sect{sec:future}, and we conclude in \sect{sec:concl}.

\section{Set-up and data acquisition}
\label{sec:setup}

The RadioAxion experiment operates underground at the Gran Sasso Laboratory. The rock overburden $\sim 1400$~m suppresses the muon and neutron fluxes by about six and three orders of magnitude, respectively, relative to the surface values. In particular, the residual muon flux is reduced to roughly one muon per square meter per hour. This strong shielding is a key ingredient of the experiment, as it renders seasonal variations of the cosmic-ray flux (with an amplitude of a few percent over the year) negligible for the dataset considered here.

The apparatus, shown in Fig.~\ref{fig:setup_am}, is installed inside the ISP8 box (surface area $\sim 10~\mathrm{m}^2$) along the corridor connecting Hall~B to the access tunnel. The temperature inside the box remains stable throughout the year, with an average of approximately $14^\circ$C and a seasonal excursion of approximately $2^\circ$C between summer and winter.

A $3''\times 3''$ NaI crystal is used to detect $\gamma$ rays from $\alpha$ decay, in particular the $59.5$~keV line, as well as X-rays from the de-excitation of $^{237}$Np (half-life $432.2$~y). The $^{241}$Am source, with an activity below $1~\mu$Ci (37 kBq), is mounted directly in front of the NaI detector. The $\alpha$ particles are fully absorbed before entering the NaI crystal. The NaI detector sees only the $\gamma$ rays and X-rays originating from the $^{241}$Am decay. 

The photomultiplier (PMT) signal is read out by an ORTEC digiBASE, a 14-pin PMT base directly coupled to the tube and integrating the high-voltage supply, preamplifier, multi-channel analyzer, and digital signal processor.
A polyethylene parallelepiped (10 cm $\times$ 10 cm $\times$ 40 cm) is shaped around the detector assembly (digiBASE, PMT, crystal and source). The parallelepiped is then 
surrounded by passive shielding consisting of $5$~cm of oxygen-free high-conductivity copper and $15$~cm of lead.

Data are acquired in list mode: each event above a $5$~keV threshold is digitized and transmitted to the acquisition PC together with its timestamp. Timing is provided by the digiBASE quartz oscillator with $1~\mu$s resolution. To mitigate frequency drifts and long-term aging of the quartz clock, each second a reference pulse derived from a rubidium frequency standard (FS725, $10$~MHz) is injected into the digiBASE. The rubidium standard has a fractional frequency stability better than $5\times 10^{-11}$ and an aging rate below $5\times 10^{-9}$ over 20 years. 

\begin{figure}[t!]
\vspace{-0.5cm}
\centerline{
\ \ \ \includegraphics[width=0.58\textwidth,angle=0]{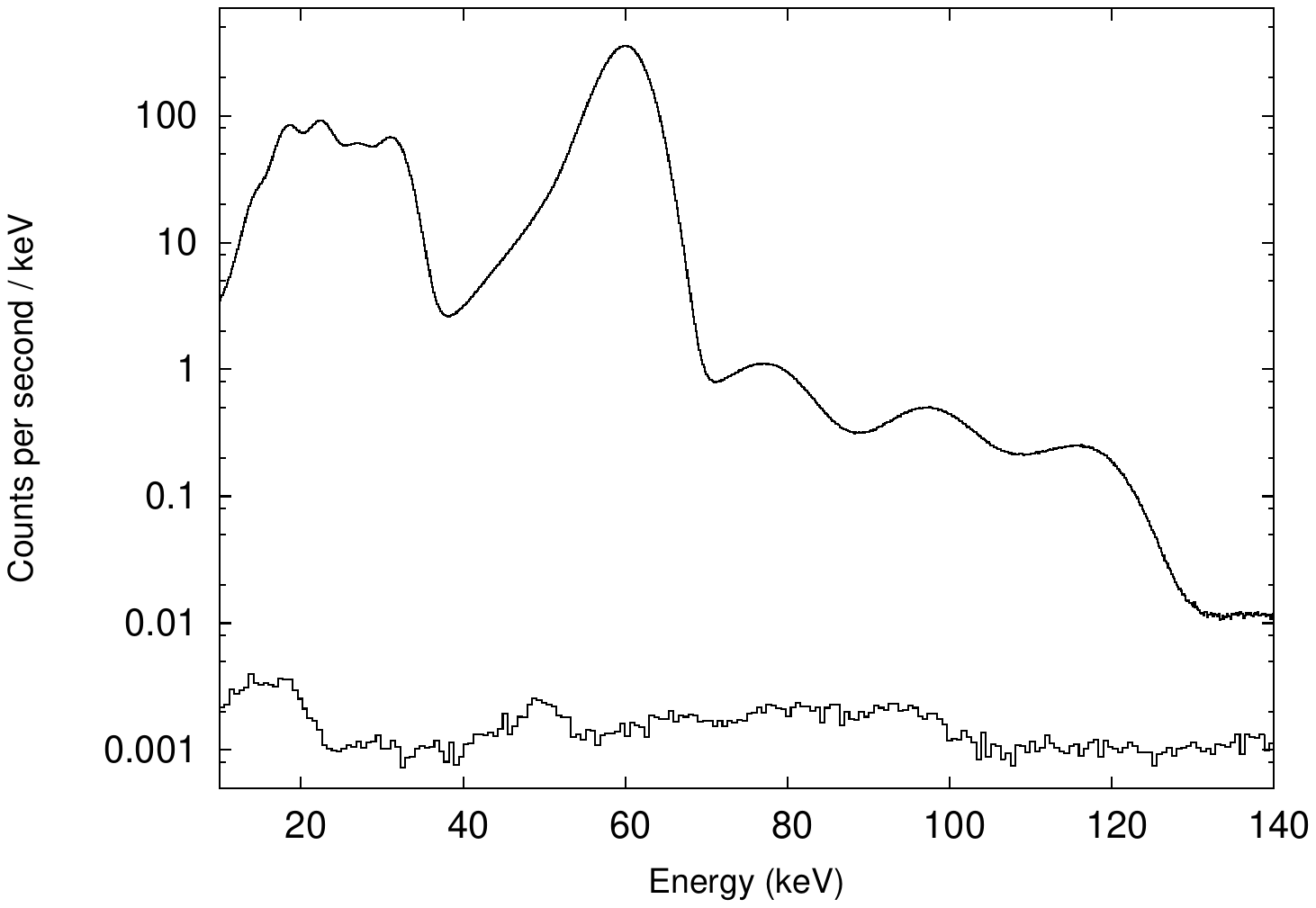}
}
\vskip -5mm
\caption{$\gamma$-spectrum (counts per second per keV) of the  $^{241}$Am source (upper curve) compared 
to the background (lower curve). 
The dominant contribution arises from the $\gamma$ at 59.5 keV. The peaks at lower energies are due to Iodine escape peak (30 keV), $^{237}$Np $\gamma$ ray (26.3 keV) 
and X-rays (17.7 and 20.7 keV). The features beyond the $59.5\,\mathrm{keV}$ line arise from: the coincidence of the $59\,\mathrm{keV}$ $\gamma$ with $^{237}$Np X-rays (yielding a peak around $75\,\mathrm{keV}$), and the $^{237}$Np $\gamma$ lines at $99$, $103$, $123$ and $125\,\mathrm{keV}$. 
The asymmetry of the 59.5 keV peak is due to the energy loss inside  the end cap of the crystal (Aluminum, 0.5 mm thick).}
\label{fig:Am_fig1}
\end{figure}

In Fig.~\ref{fig:Am_fig1} we show the energy spectrum accumulated over 24 hours, with and without the $\ce{^{241}Am}$ source. 
The dominant contribution comes from the 59.5 keV $\gamma$, producing a peak with 11$\%$ FWHM energy resolution. We point out that
our acquisition is optimized for the 
acquisition rate, due to the great number of events, not for the energy resolution (shaping time is kept at the minimum value of 0.75 $\mu$s). 

With the source in place, the total counting rate is about $4$~kHz, to be compared with a background rate of $\sim 0.2$~Hz. 
The background measured without the source is dominated by the intrinsic radioactivity of the NaI crystal, while any contribution from the external $\gamma$ rays and from the residual cosmic-ray flux are negligible.

\section{Data analysis}
\label{sec:data}


The dataset presented in this work was acquired in two configurations, enabling two complementary time-scale analyses. For the short-period analysis, targeting oscillation frequencies from $1~\mathrm{Hz}$ to $0.5~\mathrm{MHz}$, we collected 266 runs, each of one-day duration. For the long-period analysis, covering frequencies from $6.6\times10^{-7}~\mathrm{Hz}$ to $0.5~\mathrm{Hz}$ (periods from $2~\mathrm{s}$ up to $18$ days), we acquired a single continuous run lasting 69 days.

For the short-period analysis, events in the energy window $[7,130]~\mathrm{keV}$ (about $9\times10^{10}$ events in total) are grouped into consecutive $1~\mathrm{s}$ intervals, where the second is defined by the rubidium $1$-second pulse. Within each interval, timestamps recorded by the digiBASE quartz clock are corrected by linear interpolation between the rubidium timestamps marking the start and end of the interval. Slow drifts of the energy scale are corrected by tracking the position of the $59.5~\mathrm{keV}$ peak and applying a linear correction to the lower and upper edges of the analysis window.

\begin{figure*}[t!]
\centerline{
\includegraphics[width=0.49\textwidth,angle=0]{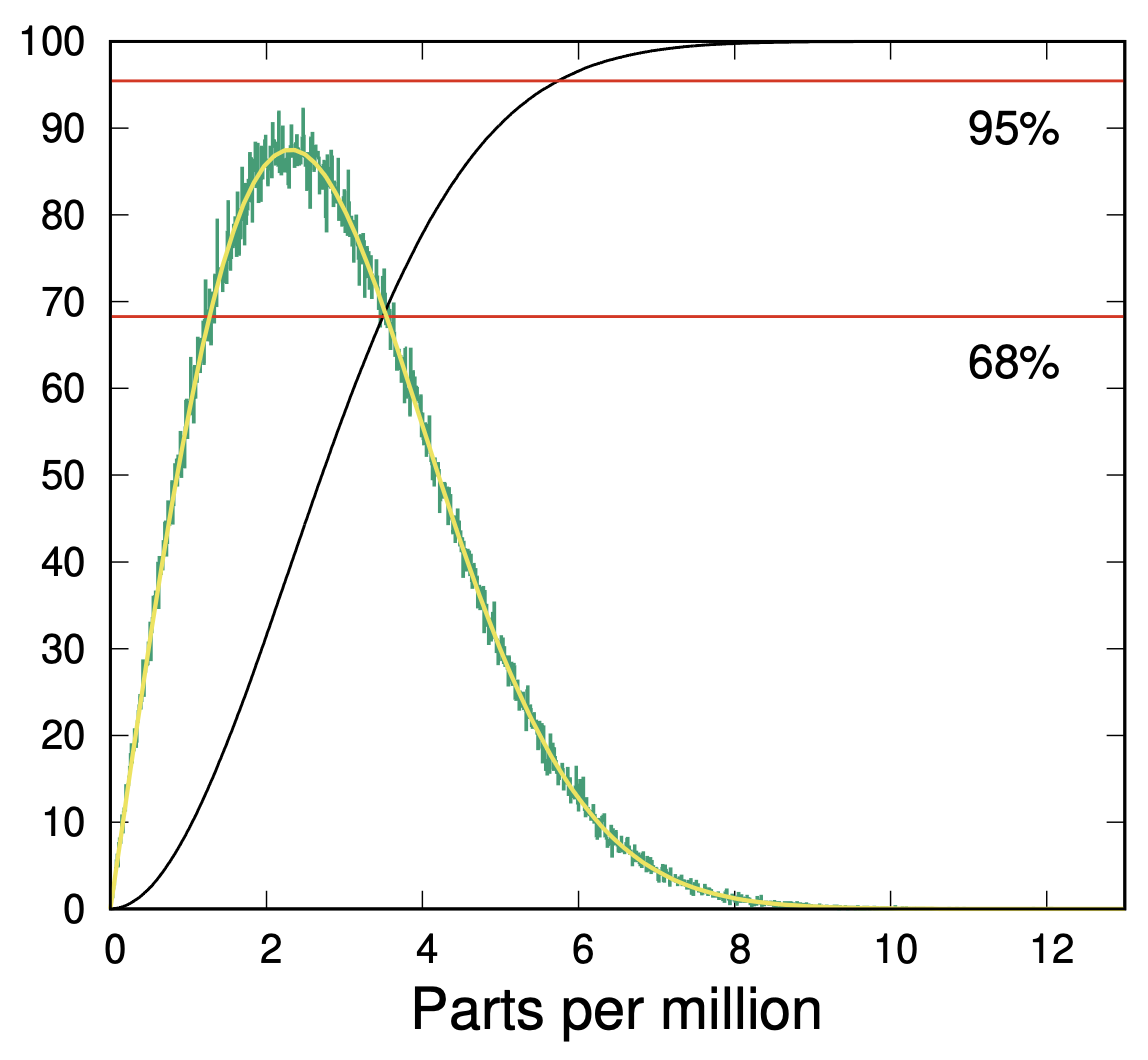}
\hspace{0.5cm}
\includegraphics[width=0.5\textwidth,angle=0]{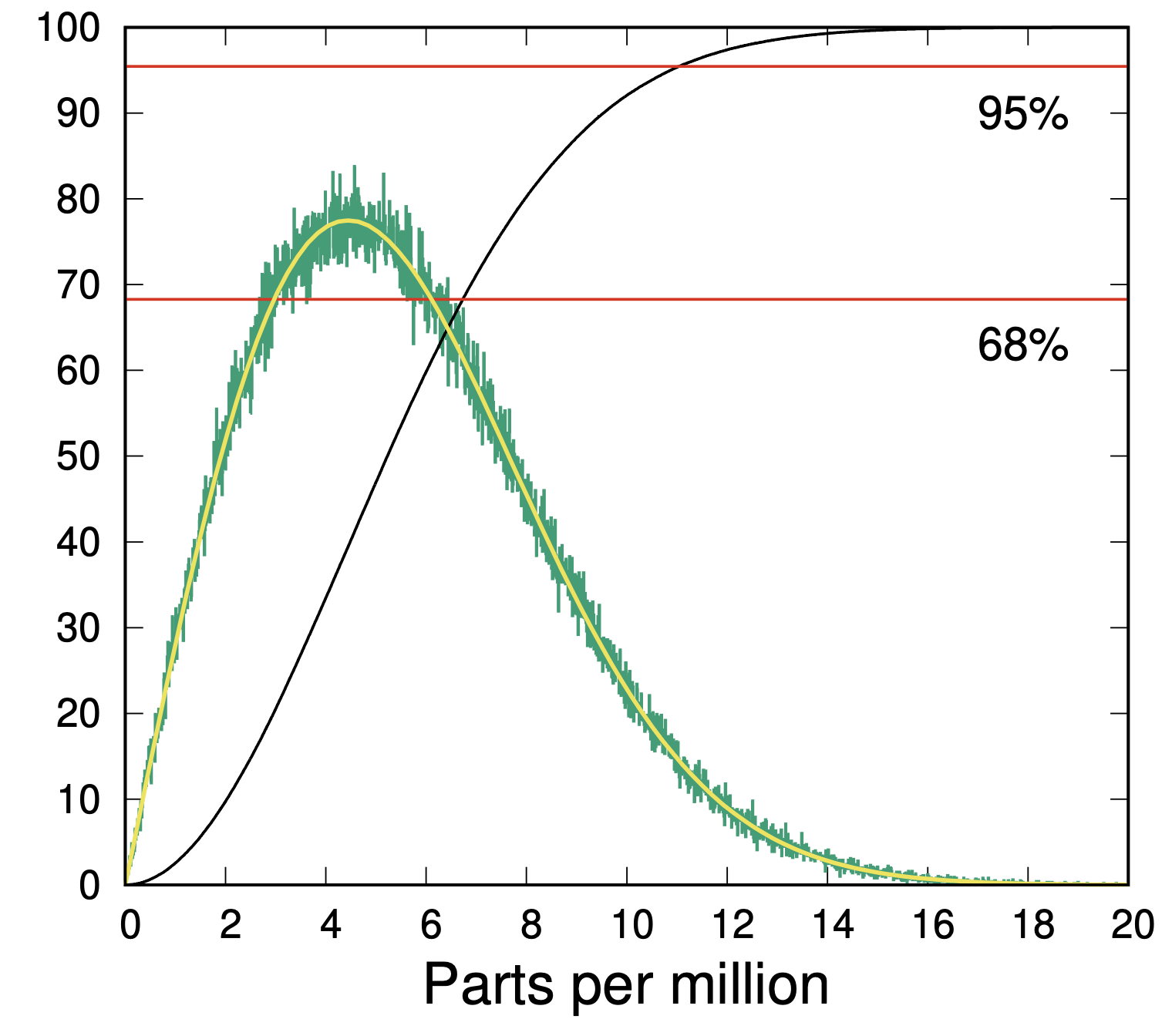}
}
\caption{\emph{Left panel:} Differential (green) ad integral (black) distributions of the normalized Fourier amplitudes for short periods ($<1\,\mathrm{s}$). The frequency range is from 1 Hz to 0.5 MHz. The two red horizontal lines are the 
68$\%$ and 
95$\%$ Confidence Levels. Differential amplitudes are in arbitrary units, integral amplitudes are in percent. 
The yellow curve shows the Rayleigh distribution expected for the modulus of the complex Fourier amplitude in the absence of an oscillatory signal, since the sine and cosine projections (with random phase) are Gaussian-distributed.
\emph{Right panel:} Same as the left panel, but for long periods ($>1\,\mathrm{s}$), corresponding to frequencies in the range $6.6\times10^{-7}\,\mathrm{Hz}$--$0.5\,\mathrm{Hz}$.
}
\label{fig:lunghi}
\end{figure*}

After these corrections, events are binned into $10^{6}$ bins of width $1~\mu\mathrm{s}$ using the fractional part of the corrected timestamp (in seconds). This yields a $10^{6}$-bin histogram with approximately $9\times10^{4}$ entries per bin. We then perform a Fourier transform to search for oscillations in the $1~\mathrm{Hz}$--$0.5~\mathrm{MHz}$ range, restricting to integer frequencies (a total of $5\times10^{5}$ modes with $1~\mathrm{Hz}$ spacing) to ensure the correct behaviour at the boundaries. 
No excess above the expected level from statistical
fluctuations is found. 
Fig.~\ref{fig:lunghi} (left) shows the differential and cumulative distributions of the normalized Fourier amplitudes. In the absence of a modulation signal, the accumulated statistics imply a $95\%$ C.L.\ limit excluding amplitudes larger than $6 \times10^{-6}$ over the $1~\mathrm{Hz}$--$0.5~\mathrm{MHz}$ frequency range.

The long-period analysis follows an analogous procedure. Events in the accepted energy window (about $2.5\times10^{10}$ in total) are binned into $1~\mathrm{s}$ intervals using the rubidium clock pulse, with $\sim 4.1\times10^{3}$ events per bin, for a total of $6{,}043{,}566$ bins. The Fourier analysis is performed at frequencies that are integer multiples of the inverse of the total corrected time duration of the measurement, again to enforce proper boundary conditions, starting from $6.6\times10^{-7}~\mathrm{Hz}$ (corresponding to a period of 18 days). In practice, robust sensitivity at long periods requires that the longest probed period does not exceed roughly one third of the total data-taking time. 
Again, no significant excess arises.
Fig.~\ref{fig:lunghi} (right) shows the differential and cumulative distributions of the normalized Fourier amplitudes. We exclude amplitudes larger than $1.1\times10^{-5}$ at the $95\%$ confidence level in the frequency range $6.6 \times10^{-7}~\mathrm{Hz}$ to about $0.5~\mathrm{Hz}$.

\begin{figure*}[t!]
\centerline{
\includegraphics[width=0.9\textwidth,angle=0]{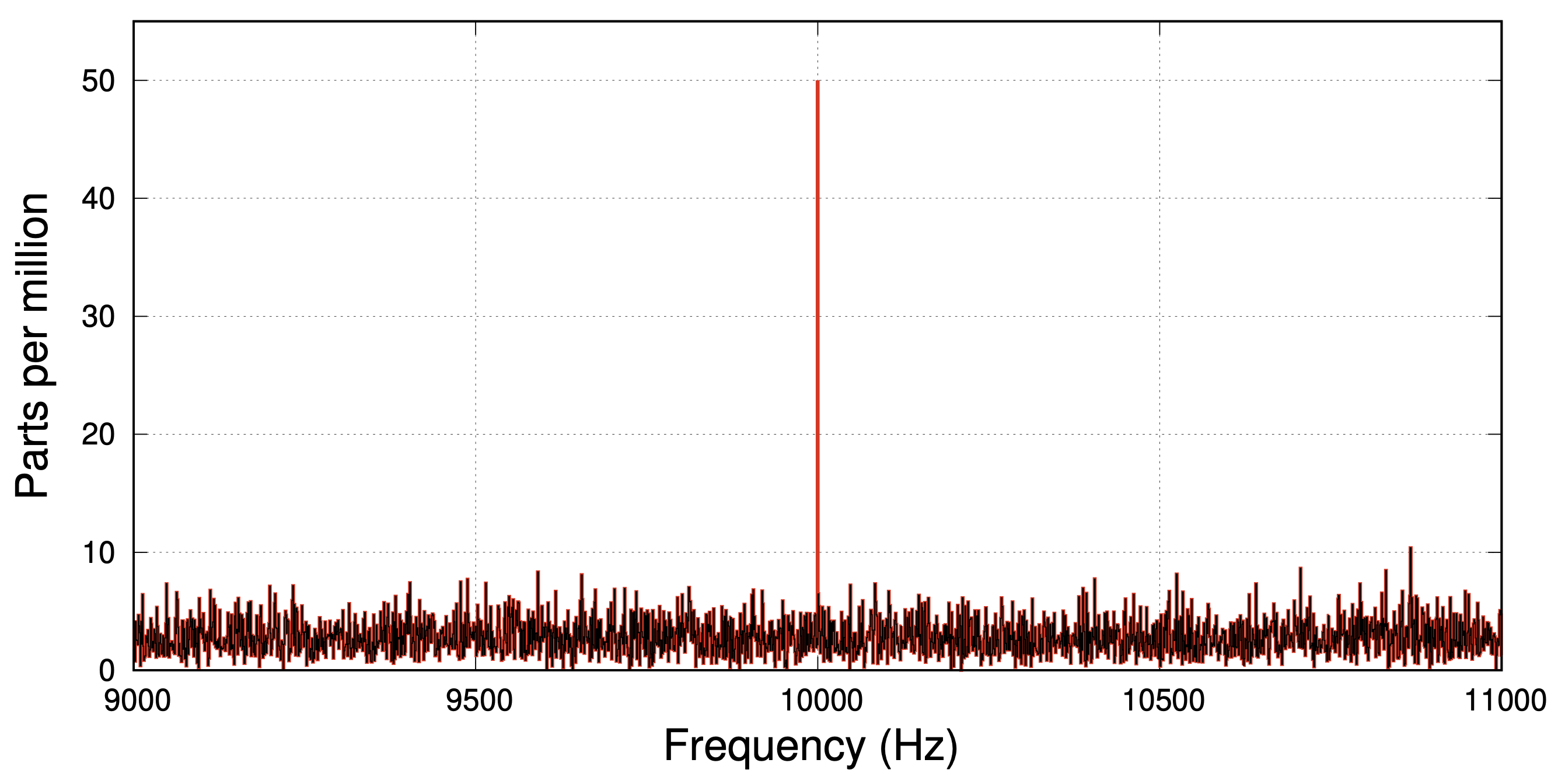}
}
\caption{Fourier spectrum in the frequency range 9--11 kHz, shown in parts per million. 
The black curve corresponds to the data alone, while the red curve shows the result after injecting a sinusoidal signal at 10 kHz with amplitude $5\times 10^{-5}$. 
The injected signal produces a narrow and clearly visible peak at the expected frequency, illustrating the sensitivity of the analysis method.
}
\label{fig:input}
\end{figure*}

To better illustrate the sensitivity of the method we injected a 10 kHz signal with $5\times10^{-5}$ amplitude.
Fig.~\ref{fig:input} shows the results of the Fourier transform in the frequency range 9-11 kHz for the data alone (black) and for the data plus the input signal (red).

\section{Axion dark matter time modulation}
\label{sec:axionDM}

We now turn to the translation of the above limits on the modulation amplitudes into constraints on axion dark matter. To this end, we briefly recall the relevant formalism from Ref.~\cite{Broggini:2024udi}.

Assuming an oscillating axion dark matter field from misalignment \cite{Dine:1982ah,Abbott:1982af,
Preskill:1982cy}, 
the time dependence of the $\theta$ angle can be approximated as
$\theta(t)=\theta_0 \cos(m_a t)$, 
with
$\theta_0=\sqrt{2\rho_{\rm DM}}/(m_a f_a)$, 
in terms of the local dark matter energy density,  
$\rho_{\rm DM} \approx 0.45 \, \text{GeV/cm}^3$.
For
a standard QCD axion, one has 
$m_a f_a = \frac{\sqrt{m_u m_d}}{m_u + m_d} m_\pi f_\pi \approx (76 \ \text{MeV})^2$, 
corresponding to $\theta_0 \approx 5.5 \times 10^{-19}$. In the following, we will treat $m_a$ and $f_a$ as independent parameters and discuss the limits from $\alpha$-decay in the $(m_a, 1/f_a)$ plane. 
QCD axion models that modify the standard $m_a$--$f_a$ relation (in particular, scenarios in which the axion mass is parametrically suppressed at fixed $f_a$ by a symmetry principle) are discussed e.g.~in \cite{Hook:2018jle,DiLuzio:2021pxd,DiLuzio:2021gos,Banerjee:2022wzk,Banerjee:2025zcd}. 

Following Ref.~\cite{Broggini:2024udi}, we introduce the 
observable
\beq 
\label{eq:Itdef}
I(t) \equiv \frac{T^{-1}_{1/2}(\theta(t)) - \langle T^{-1}_{1/2} \rangle }{\langle T^{-1}_{1/2} \rangle } \, , 
\eeq
where $T^{-1}_{1/2}(\theta(t))$ denotes the inverse half-life of $\ce{^{241}Am}$ as a function of $\theta(t)$, and $\langle T^{-1}_{1/2}\rangle$ is its time average. 
Given that the  
$\theta$-dependence of $T_{1/2}(\theta)$ 
is analytic in $\theta^2$ \cite{Broggini:2024udi}, 
it 
admits the Taylor expansion
$T_{1/2}(\theta) \approx T_{1/2}(0) + \mathring{T}_{1/2}(0) \theta^2$,  
where we introduced the quadratic slope, $\mathring{f} \equiv df/d\theta^2$. 
Since $\theta^2 \ll 1$, the truncated small-$\theta$ expansion provides an excellent approximation to the full $\theta$-dependence.
Using $\langle \cos^2 (m_a t) \rangle = 1/2$ 
and expanding at the first non-trivial order in $\theta_0$, 
one finds 
\begin{align} 
\label{eq:Itpred}
I(t) &\approx - \frac{1}{2} \frac{\mathring{T}_{1/2}(0)}{T_{1/2}(0)} \theta_0^2 \cos(2 m_a t) \nonumber \\
&= - 4.3 \times 10^{-6} \cos(2 m_a t) \left( \frac{\rho_{\rm DM}}{0.45 \, \text{GeV/cm$^3$} } \right) \nonumber \\
&\times \left( \frac{10^{-16} \, \text{eV}}{m_a} \right)^2  \left( \frac{10^{8}  \, \text{GeV}}{f_a} \right)^2
\, ,  
\end{align}  
where  
$\mathring{T}_{1/2}(0) / T_{1/2}(0) \approx 125$
for $\ce{^{241}Am}$ \cite{Broggini:2024udi}.

The theoretical prediction in \eq{eq:Itpred} can be compared with the experimental estimator
$I_{\rm exp}(t)\equiv\bigl(N(t)-\langle N\rangle\bigr)/\langle N\rangle$,
where $N(t)$ is the observed number of events in a given time interval and $\langle N\rangle$ its expected value. 
Note that $I_{\rm exp}$ corresponds to the variable on the $x$-axis of Fig.~\ref{fig:lunghi}. 

Potential sources of systematic errors include the detection of $\gamma$-rays and their time-stamping. 
The former is mitigated by operating the NaI detector well-below the radiation damage threshold 
and by the reduced background in the underground environment. 
The latter is handled thanks to the precision of a Rb atomic clock.\footnote{The axion dark matter background also induces a fractional shift of the Rb hyperfine reference frequency,
$(\delta f /f)_{\rm Rb} \approx 4 \times 10^{-10} ( 10^{-16} \, \text{eV} / m_a )^2  ( 10^{8}  \, \text{GeV} / f_a )^2$
through the $\theta$ dependence of the proton mass and of nuclear $g$-factors~\cite{Zhang:2022ewz}. In the mass/coupling range relevant for our bounds this corresponds to an absolute shift of order $\delta f\sim 1\,\mathrm{Hz}$ for the standard Rb frequency $f_{\rm Rb}\approx 6.8\,\mathrm{GHz}$. Equivalently, the induced timing error over an interval of length $t$ is
$\delta t \approx t\,(\delta f/f)_{\rm Rb}\sim 10^{-10}\,t$,
so that for $t=1\,\mathrm{s}$ one finds $\delta t\sim 10^{-10}\,\mathrm{s}$. This is far below the level relevant for our analysis, which is sensitive to much larger modulations in the event-time distribution (at most at the $10^{-6}$ level). Therefore axion-induced shifts of the Rb reference do not affect, nor can they cancel, the signal we search for.}
Hence, we expect our uncertainties to be statistically dominated in the current setup, with the number of events being the limiting factor, i.e.~the $^{241}$Am source activity.

\begin{figure*}[t!]
\centerline{
\includegraphics[width=0.9\textwidth,angle=0]{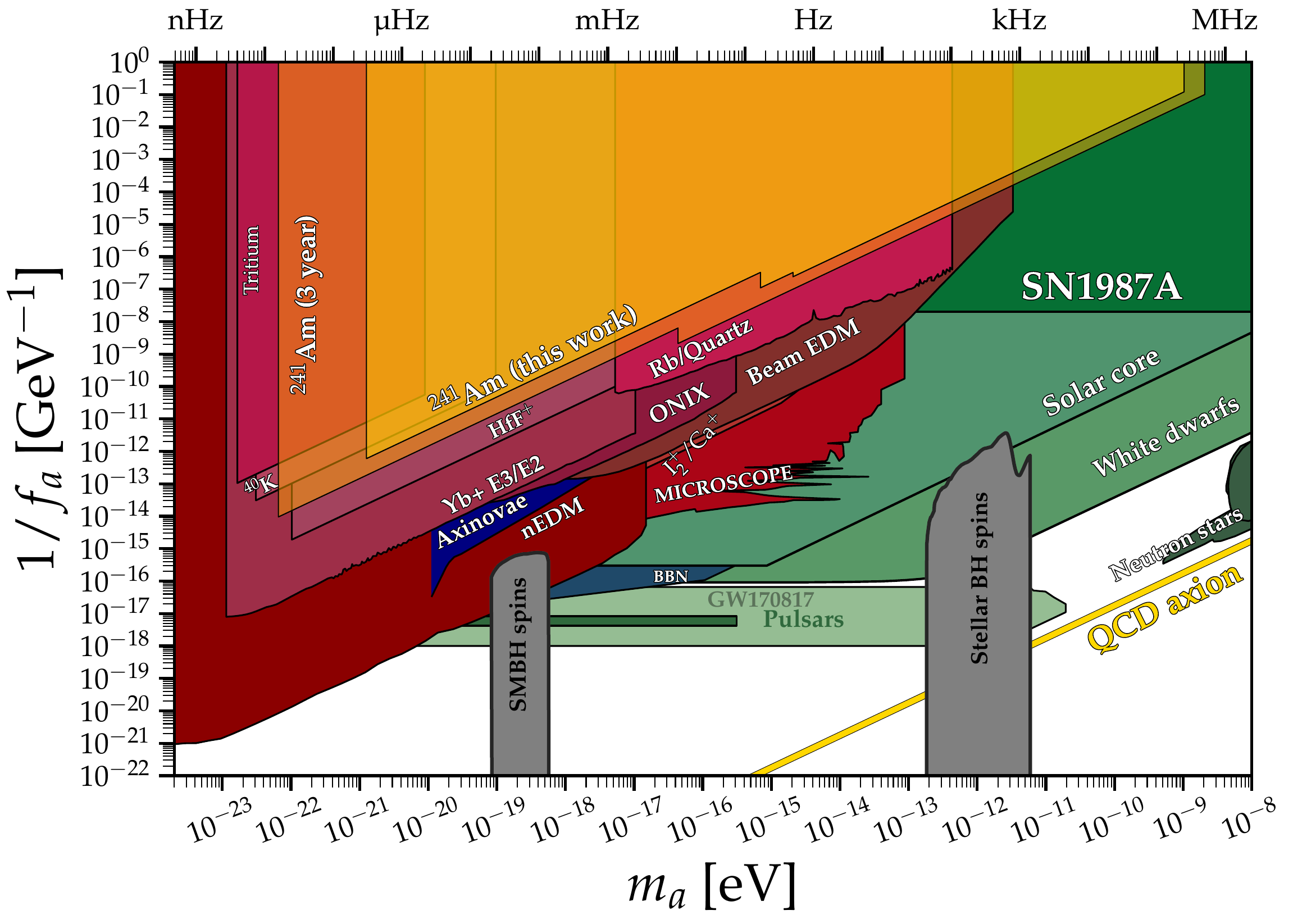}
}
\caption{Constraints on the axion decay constant as a function of the axion mass. 
The exclusion limits derived in this work are shown in the yellow region, together with the projected sensitivity of a 3-year data taking (orange area). 
The result is displayed as a continuous band for readability, although the analysis is performed at a discrete set of frequencies, as described in \sect{sec:data}.
Limits from other laboratory experiments and astrophysics/cosmology are shown as well for comparison
(see text for details). Figure adapted from \cite{AxionLimits}. 
}
\label{fig:bounds}
\end{figure*}

An important remark concerns the coherence time of the axion dark matter signal, $\tau_c \approx 2\pi/(m_a v^2)$, with $v\approx 10^{-3}$ the virial velocity in natural units. As discussed in Ref.~\cite{Centers:2019dyn}, if the measurement time $T_{\rm exp}$ is much shorter than $\tau_c$, the experiment does not sample the full stochastic distribution of axion field amplitudes. In this regime it is therefore not appropriate to infer limits assuming a fixed value of $\theta_0$ determined solely by the averaged local dark matter density. Following Ref.~\cite{Centers:2019dyn}, for $T_{\rm exp}\lesssim \tau_c$ we account for this effect by rescaling the deterministic $2\sigma$ bound as $(1/f_a)^{\rm stoch}_{2\sigma}\approx 3.0\,(1/f_a)^{\rm det}_{2\sigma}$.

For the short-period dataset we take $T_{\rm exp}^{(1)}\approx 2.3\times 10^{7}$~s, therefore the stochastic correction applies for axion masses smaller than $2\pi\times 10^{6}/T_{\rm exp}^{(1)}\approx 1.8\times 10^{-16}$~eV, or equivalently for frequencies $\lesssim 0.044$~Hz. 
Since we report short-period bounds only for frequencies above $1\,\mathrm{Hz}$, this correction does not enter the short-period analysis.
For the long-period dataset, $T_{\rm exp}^{(2)}\approx 6.0\times 10^{6}$~s, which implies that the correction applies to axion masses smaller than $2\pi\times 10^{6}/T_{\rm exp}^{(2)}\approx 6.9\times 10^{-16}$~eV, or equivalently for frequencies $\lesssim 0.17$~Hz. 
In this case the correction is relevant over most of the frequency interval probed by the long-period analysis.

The exclusion limits derived in this work are shown by the yellow region in \fig{fig:bounds}. 
The small step around $0.5$ Hz marks the transition between the two datasets. Note that the upper $x$-axis indicates the Compton frequency associated with the axion mass, $m_a/(2\pi)$, which is \emph{half} the expected signal frequency (see Eq.~\eqref{eq:Itpred}). 
The other step, at $0.17$ Hz, arises from the stochasticity correction discussed above. 
The projected sensitivity 
of a future three-year RadioAxion data-taking period 
(with improved setup described in \sect{sec:future})
is indicated by the orange region in the same figure, 
with the step at $0.01$ Hz denoting the stochasticity correction. 
Previous limits from the time-modulation of radioisotope 
decays of $\ce{^{40}K}$ \cite{Alda:2024xxa} 
(corrected by the stochasticity factor)
and Tritium 
\cite{Zhang:2023lem} decay are indicated in red. 

For comparison, we also display laboratory limits from 
EDM searches~\cite{Abel:2017rtm,Roussy:2020ily,JEDI:2022hxa,Schulthess:2022pbp}, 
radio-frequency atomic transitions~\cite{Zhang:2022ewz},  and molecular clocks~\cite{Madge:2024aot}, 
as well the model-independent SN 1987A bound~\cite{Lucente:2022vuo,Springmann:2024ret},
finite-density-induced bounds from the solar core and white dwarfs~\cite{Hook:2017psm,Balkin:2022qer}, 
neutron stars cooling~\cite{Gomez-Banon:2024oux,Kumamoto:2024wjd},
gravitational waves~\cite{Hook:2017psm,Zhang:2021mks},
black hole superradiance~\cite{Cardoso:2018tly,Baryakhtar:2020gao,Unal:2020jiy,Hoof:2024quk,Witte:2024drg}, 
and other cosmological probes~\cite{Blum:2014vsa,Fox:2023xgx}.

As is apparent from \fig{fig:bounds}, the present RadioAxion limits are weaker than the leading existing constraints over most of the parameter space. 
They should therefore be viewed primarily as a proof of principle for a complementary laboratory strategy based on continuous, underground monitoring of nuclear decay rates. 
As discussed in \sect{sec:future}, the sensitivity can be improved with larger statistics, higher source rate, and longer exposure. 
The ultimate reach is expected to be limited mainly by the achievable counting statistics and by the fact that decay-rate modulations arise from $\theta^2$ effects, unlike EDM searches which are linear in $\theta$. 
A distinctive feature of the method is its access to high oscillation frequencies, including the kHz range and above, which are challenging for other laboratory techniques, although this region is already disfavoured by astrophysical bounds.

\section{Future developments}
\label{sec:future}

The second phase of the RadioAxion experiment will run for three years, with the goal of improving statistics and extending the sensitivity to longer oscillation periods through an upgraded setup.

A $1''\times 1''$ CeBr$_3$ crystal will be used to detect the $\gamma$ rays associated with $\alpha$ decays. CeBr$_3$ was chosen for its much shorter scintillation decay time and its reduced susceptibility to radiation damage compared to NaI. The readout electronics will be upgraded from the digiBASE to the digiBASE-E, providing a timing resolution of 160~ns. In addition, ten $\ce{^{241}Am}$ sources, each with an activity below $1~\mu$Ci, will be mounted in a plastic holder surrounding the CeBr$_3$ crystal, increasing the event rate by about one order of magnitude, up to 40 kHz.

To improve long-term stability, two heating tapes will be used to regulate the temperature of the electronics directly attached to the photomultiplier, keeping it constant within $0.1^\circ$C. We verified that temperature changes at the level of $2^\circ$C, as observed inside the ISP8 box between summer and winter, are sufficient to induce sizable variations in the shaping-time constant of the digiBASE amplifier by affecting the capacitor and resistance values. This in turn would modulate the dead time value generating a fake periodic signal.

Finally, the fully shielded detector assembly, including the copper and lead layers, will be enclosed in a Plexiglass box to further isolate it from the environment. In particular, this would protect the experiment from the radon inside the laboratory, whose amount fluctuates in time as a function of pressure and temperature. If necessary, it will be possible to keep a small overpressure inside the box by feeding it with pure nitrogen gas. 

The apparatus is now ready for commissioning, and we plan to start data taking in spring 2026.
We expect to reach a $2\sigma$ error of $2/\sqrt{10\times4000/s\times3\text{ year}}\approx1\times10^{-6}$ on $I_\text{exp}$ after three years of operation. Requiring the observation of at least three full oscillations, we can test periods up to one year. The corresponding sensitivity in the axion parameter space is shown by the orange area in Fig.~\ref{fig:bounds}.

\section{Conclusions}
\label{sec:concl}

RadioAxion is an underground experiment that searches for axion dark matter through periodic modulations of radioactive decays. In this work we presented results from the first RadioAxion data set, based on monitoring the $\alpha$ decay of $\ce{^{241}Am}$ via the $59.5$~keV $\gamma$ line with a NaI detector installed at the Gran Sasso Laboratory, where cosmic-ray-induced systematics are strongly suppressed.

We reported the measured energy spectra and performed a time-series analysis in two complementary frequency intervals, using 266 one-day runs to probe $1~\text{Hz}$--$0.5~\text{MHz}$ and a continuous 69-day run to probe $6.6 \times10^{-7}~\text{Hz}$--$0.5~\text{Hz}$. No evidence for a periodic modulation was observed. We therefore set $2\sigma$ upper limits on the modulation amplitude at the level of $6 \times10^{-6}$ at short periods and $1.1\times10^{-5}$ at long periods.

Using the $\theta$-dependent $\alpha$-decay framework of Ref.~\cite{Broggini:2024udi}, these bounds translate into new exclusion limits on the axion decay constant over axion masses spanning $10^{-21}$ to $10^{-9}$~eV, as shown in \fig{fig:bounds}. 
Although these first limits are weaker than the leading existing constraints, they establish the sensitivity of an independent laboratory approach based on nuclear decay-rate modulation.
Finally, we outlined the upgraded three-year phase of RadioAxion, which will increase the event rate and improve timing and thermal stability, with the projected sensitivity also reported in \fig{fig:bounds}.


\section*{Acknowledgments}

We thank Jorge Alda Gallo and Stefano Rigolin for discussions during the early phase of the data analysis.
LDL is supported by the European Union -- Next Generation EU and
by the Italian Ministry of University and Research (MUR) 
via the PRIN 2022 project n.~2022K4B58X -- AxionOrigins.
The work of CT has received funding from the French ANR, under contracts ANR-19-CE31-0016 (`GammaRare') and ANR-23-CE31-0018 (`InvISYble'), that he gratefully acknowledges.



\bibliographystyle{utphys}
\bibliography{bibliography.bib}


\end{document}